\documentstyle[aps,prb,twocolumn,epsf,rotate]{revtex}

\newcommand{\beq}{\begin{equation}}
\newcommand{\eeq}{\end{equation}}
\newcommand{\rbq}{{\bar \rho}_{\bf q}}

\newcommand{\qq}{{\bf q}}
\newcommand{\pp}{{\bf p}}

\begin{document}
%% The following two lines should be there when using 'twocolumn'.
\twocolumn[\hsize\textwidth\columnwidth\hsize\csname
@twocolumnfalse\endcsname

\title{Liouvillian Approach to the Integer Quantum Hall Effect Transition}

\draft

\author{Jairo Sinova$^{1,2}$, V. Meden$^{1,3}$, and S.M. Girvin$^{1}$}
\address{$^{1}$Department of Physics,
Indiana University, Bloomington, Indiana 47405-7105}
\address{$^{2}$ Department of Physics and Astronomy, University of Tennessee, 
Knoxville, Tennessee 37996-1200}
\address{$^{3}$Institut f\"ur Theoretische Physik, 
Universit\"at G\"ottingen, Bunsenstr.\ 9, D-37073 G\"ottingen,
Germany}
\date{\today}
\maketitle

\begin{abstract}
We present a novel approach to the  
localization-delocalization 
transition in  the integer quantum Hall effect. The Hamiltonian projected onto 
the lowest Landau level can be written in terms of the %V0205
projected density operators alone. This and the closed set of   %J0203
commutation relations between the projected densities leads to simple
equations for the time evolution of the density operators. These
equations can  be used to map the problem of calculating the 
disorder averaged and energetically unconstrained density-density
correlation function to the problem of calculating the
one-particle density of states of a dynamical system with 
a novel action. At the self-consistent mean-field level, %V0205
this approach yields normal diffusion and a finite longitudinal 
conductivity. 
While we have not been able to go beyond the saddle point
approximation analytically, we show numerically that the 
critical localization exponent can be extracted 
from the energetically integrated correlation function yielding    %J0206
$\nu=2.33 \pm 0.05$ in excellent agreement with previous 
finite-size scaling studies.
\end{abstract}

 %J0203

%% The following line should be there when using 'twocolumn'.
\vskip2pc]

%%%%%%%%%%%%%%%%%%%%%%%%%%%%%%%%%%%%%%%%%%%%%%%%%%%%%%%%%%%%%%%%%%%%

\section{Introduction}

The metal-insulator transition in the integer quantum Hall
effect (IQHE) is a reentrant zero temperature 
quantum phase transition 
in which the sample goes from an insulating 
phase with longitudinal conductivity $\sigma_{\rm xx}=0$ to 
another insulating phase by crossing a conducting critical 
point  ($\sigma_{\rm xx}\ne0$) as the magnetic field is varied. 
The critical point occurs between the plateaus of the 
Hall conductivity $\sigma_{\rm xy}$
and corresponds to the instance when the Fermi energy is 
at a critical energy located in the middle of one of the 
disorder broadened Landau levels.\cite{Hajdu book,Huckestein RMP}   

In general, the disorder induced metal-insulator transition 
is a transition in the
nature of the states (whether they are localized
or delocalized) at the Fermi energy 
and it does not manifest itself in the density of states
which remains smooth across the mobility edge. 
According to the one-parameter theory of 
scaling, the states of a two-dimensional noninteracting electron gas
are all localized in the presence of arbitrary 
weak disorder.\cite{group_of_four}
In the IQHE however, the presence of the  
strong magnetic field pointing perpendicular to the plane  %J0203
drastically changes the nature of the states near the middle of 
the Landau bands.
In the noninteracting picture of the IQHE these states are characterized
by a localization length 
\begin{eqnarray}
\xi(E) \sim \xi_0 \left| \frac{E-E_c^{i}}{E_0} \right|^{-\nu}
\label{xidef}
\end{eqnarray} 
which determines the extent to which the eigenstates of energy 
$E$ are delocalized. Here $\xi_0$ denotes 
a characteristic length scale of the system, e.g.\ the magnetic length 
$\ell$ (see below) and $E_0$ a characteristic energy scale, 
e.g.\ the bandwidth or disorder strength.
The critical energy $E_c^{i}$ is located in the middle of the 
$i$-th Landau band and, in an infinite size system, 
it is the only energy at which the one-particle 
eigenstates are delocalized within this Landau band. 
As the Fermi energy (or magnetic field) is varied,
the conductivity $\sigma_{\rm xx}$     %S0209
will change according to the nature of the states at that energy
and sharp peaks in the longitudinal conductivity will be observed.
%S0209

When studying the IQHE the interaction between the electrons 
is usually ignored and only 
the disorder is considered to be 
responsible for the localization of the single particle 
states.
This assumption must be checked by comparing the predictions
of the noninteracting theory to experimental 
results\cite{Hajdu book,Huckestein RMP,HPWei} and the outcome
of numerical calculations which include the 
interactions.\cite{Yang,ZQW,Huckeandback}
The universal localization exponent $\nu=2.34 \pm 0.04$
numerically obtained within a noninteracting 
theory\cite{Huckesteinkramer,Mieck,huo_and_bhat} is in excellent
agreement with experimental measurements of 
$\nu$,\cite{Hajdu book,Huckestein RMP,HPWei}  but it
remains a mystery why the strong interactions, which do affect 
the dynamical exponent $z$, 
does not seem to affect $\nu$.\cite{Yang,ZQW,Huckeandback,Steve_Chalker} 
Here we adopt the noninteracting picture. 
We furthermore assume a strong magnetic 
field and a Zeeman splitting, which is much larger than the width of 
each disorder broadened Landau level. We can then focus on the transition 
within the lowest Landau level (LLL) and neglect the spin degree of
freedom of the electrons.

It has been shown numerically that for 
a finite system delocalized one-particle wave functions near $E_c$
show multifractal properties characterized by a set of
generalized fractal dimensions ${D}_q$.\cite{Hajdu book,Chalker,multifractal}
Also, dynamical studies have shown
anomalous slow diffusion of wave packets constructed from 
these multifractal states.\cite{Hucke}
Diffusion can be studied using the spectral function of the 
disorder averaged retarded  %J0203
density-density correlation function.\cite{Forster} 
For the problem considered here %J0203 
the spectral function is given by
(after dividing by $\pi \hbar \omega$)\cite{Chalker}  %V0210   
\begin{eqnarray}
\bar S(r,\omega;E)&\equiv& 
\Bigl< \! \Bigl<  
\sum_{i,j} \delta(E- \hbar \omega/2 - E_i)
\delta(E+ \hbar \omega/2 - E_j) \nonumber
\\* &&
\times 
\psi_i(0)\psi_i^*({\bf r})\psi_j({\bf r})\psi_j^*(0)
\Bigr> \! \Bigr>  \, .
\label{Sdef}
\end{eqnarray}
Here the $\psi_{i}({\bf r})$ denote one-particle 
eigenfunctions and
$E_{i}$ the respective eigenenergies 
for an electron of a two-dimensional 
spinless electron gas  which is subject to a perpendicular 
magnetic field and a disorder potential.
$\langle \! \langle \ldots \rangle \! \rangle$ indicates the ensemble
average over the disorder. After taking the disorder average,  %J0206
translational invariance is restored and 
$\bar S$ only depends on the distance 
$r \equiv |{\bf r}|$ from the origin of the
plane. 
Assuming that the eigenstates which contribute in Eq.\ (\ref{Sdef})
for $E \approx E_c$ are of multifractal character,
it has been argued that $\bar S$ decays algebraically\cite{Hajdu book,Wegner2}
\begin{equation}
\bar S(r,\omega\rightarrow 0;E\rightarrow E_c)\sim 
\left(\frac{r}{\xi(E)}\right)^{-\eta} \,\,,
\end{equation}
for $\xi_0 \ll r  \ll  \xi(E)$. 
The anomalous diffusion exponent $\eta$ is related
to the generalized fractal dimension via 
${D}_2=2-\eta$.\cite{Hajdu book,multifractal}

Assuming a generalized nonlocal (in time and space) relation between
the current and the gradient of the density and using the continuity
equation, the spectral function in momentum space $S(q,\omega;E)$
at small $q \equiv |{\bf q}|$ and $\omega$,
can be rewritten in terms of a generalized diffusion 
``coefficient'' $D(q,\omega)$ for $E \approx E_c $\cite{Forster}
\begin{equation}
S(q,\omega;E)=\frac{\rho(E)}{\pi}
\frac{\hbar q^2 D(q,\omega)}{[\hbar \omega]^2+
[\hbar q^2  D(q,\omega)]^2}\,\,,
\label{sind}
\end{equation}
where $\rho(E)$ is the density of states per unit area. 
In the limit of $\omega,q\rightarrow 0$ and for large
enough system sizes, $D(q,\omega)$ is
only a function of $qL_\omega$, where\cite{Chalker}
\begin{eqnarray}
L_\omega \equiv [\rho(E_c)\hbar\omega]^{-1/2} \, .
\label{Lomegadef} 
\end{eqnarray}
Through numerical diagonalization   
and using Eq.\ (\ref{sind}) Chalker and  Daniell\cite{Chalker} 
have shown that $D(q,\omega)$ approaches a constant $D_0$ 
for small  $qL_\omega$. The precise value of $D_0$ is 
important since the longitudinal conductivity at the critical 
point is given by the Einstein relation $\sigma_{\rm xx} = e^2 \rho(E_c) D_0$ 
and is expected to be universal. The $qL_\omega\rightarrow 0$ limit of  %J0203
$D(q,\omega)$ has later been reinvestigated in an extended 
numerical study.\cite{huoandbhatt2} 
For  $ qL_\omega \gg 1$,   %V0205
but still in the limit of $q,\omega \to 0$, $D(q,\omega)$ 
decays as\cite{Chalker} 
\begin{equation}
D(q,\omega)\propto D_0 (q L_\omega)^{-\eta} \, .
\end{equation}
For the anomalous diffusion exponent $\eta$ Chalker and  Daniell
obtain the numerical value $\eta=0.38 \pm 0.04$, indicating
that the delocalized states near the critical energy
indeed have multifractal properties. This value for $\eta$ has later
been confirmed in other numerical studies.\cite{Hucke,Tobias}
For energies $E$ away from $E_c$, $S(q,\omega;E)$ vanishes in the
small $q$ and $\omega$ limit independent of the order in which the
limits are taken due to exponential localization of the states.

Most of the progress in the theoretical understanding
of the localization-delocalization transition considered here 
has been through numerical 
calculations.\cite{Huckestein RMP,Chalker,Hucke}  
Although a field theory has been proposed some time ago by Pruisken and   %J0206
co-workers,\cite{Pruisken} up to now no quantitative results such as the 
critical exponents of the transition have been obtained within this 
description. More recent studies\cite{Zirnbauer,Bhaseen} have introduced
alternative field theories. Within the framework of these theories   
it might in the future be possible to analytically determine critical
exponents as has been recently successfully achieved for the SU(2)
version of the network model.\cite{Grizberg}  %S0209
In this paper we present a novel approach to the transition
which may prove more tractable. 
Although thus far we have not been able to analytically
calculate the spectral function Eq.\
(\ref{Sdef}) beyond the self-consistent Born 
approximation, we have  numerically verified 
the possibility of obtaining the critical exponent $\nu$ using 
this approach.

We start by defining the density correlation function at zero temperature as
\begin{equation}
\tilde\Pi(q,t;E)\equiv 
-\frac{i\theta(t)}{N \hbar \ell^2} 
\left< \! \left< {\rm Tr}\{
\bar{\rho}_\qq(t)\bar{\rho}_{-\qq}(0)\delta(E-H)\}\right> \! \right> 
\,\,,
\label{eqn1}
\end{equation}
with the one-particle Hamiltonian $H=H_0+H_D$. Here $H_0$  
denotes the kinetic energy of a spinless 
electron moving in the plane    %V0205
in the presence of a perpendicular magnetic field 
and $H_D$ is the potential energy for a fixed  
realization of the disorder potential $V({\bf r})$.  
$\ell$ is the magnetic length given by $\ell^2= \hbar c/(eB)$, where 
$B$ is the strength of the magnetic field, and  
$N=L^2/(2 \pi \ell^2)$ is the number of states in the LLL. %J0203
We consider a square sample of area $L^2$. %J0203
By projecting the one-particle 
density operator $\rho_\qq \equiv \exp{(-i{\bf q} \cdot {\bf r})}$ 
onto the LLL, denoting the projected density by $\bar\rho_\qq$ (see
Sec.\ II), and taking the one-particle trace ${\rm Tr}$ over the states
in the LLL, we restrict our considerations to the transition in the
LLL. 
It will turn out that the equation of motion for
the density operators restricted to the LLL can be
solved formally in this case.  %J0214
In the small $\omega$ limit we have
\begin{eqnarray}
S(q,\omega;E) = - \frac{1}{2 \pi^2} {\rm Im} \,  \Pi(q,\omega;E) \, .
\label{relation}
\end{eqnarray}
Instead of dealing with $\tilde\Pi(q,t;E \approx E_c)$ directly,
we will integrate $\tilde\Pi(q,t;E)$ over {\it all} energies $E$ 
and focus our attention on 
\begin{equation}
\tilde\Pi(q,t)\equiv -i\frac{\theta(t)}{N \hbar \ell^2}
\langle \! \langle{\rm Tr}\{
\bar{\rho}_\qq(t)\bar{\rho}_{-\qq}(0)\}\rangle \! \rangle\,\,.
\label{pidef}
\end{equation}
Since the localization length only diverges
at $E_c$, the energetically unconstrained diffusion 
problem considered by investigating ${\rm Im} \, \Pi(q,\omega)$  
still contains useful information about critical exponents. 
For instance, let us suppose that at time $t=0$ 
we create a wave packet localized at the origin %J0203
constructed from {\it all} the states of the system 
(localized as well as delocalized states).
For large $t$  only ``delocalized'' states with $\xi(E)/r>1$ 
can contribute to the probability amplitude of the wave packet at 
a distance $r/\xi_0 \gg 1$ far away from the origin. 
This implies that in the limit of small  $q \xi_0$ only
states with $q \xi(E)>1$ and thus %[see Eq.\ (\ref{xidef})]  %J0203
$|E-E_c| < E_0 (q \xi_0)^{1/\nu}$ contribute to the right hand side 
of Eq.\ (\ref{pidef}). 
Hence for $q \xi_0 \to 0$ only a fraction 
$\sim(q \xi_0 )^{1/\nu}$  of the states in the LLL contributes %V0210 
and we expect from Eq.\ (\ref{sind}), that for small $qL_\omega$
\begin{equation}
- \hbar \omega \ell^2 \; {\rm Im} \, \Pi (q,\omega) \propto
(q \xi_0)^{1/\nu} \frac{(D_0 q^2/\omega)}{1+(D_0 q^2/\omega)^2}\,\,,
\label{argument}
\end{equation}
where the diffusion parameter is a constant $D_0$.
The above argument, which we confirm 
numerically in Sec.\ V, gives a strong indication 
that some useful information about the quantum phase  
transition can be extracted from $\Pi(q,\omega)$.
We again emphasize that this is so because the 
delocalization only occurs at a single critical 
energy, a characteristic unique to the IQHE
where the extended states have zero measure in the energy
spectrum. We also point out the importance of the order of limits in
obtaining Eq.\ (\ref{argument}). 
The limit of $q$,$\omega\rightarrow 0$
is taken by having $q$ approach zero faster than $\omega$ so as
to obtain a finite diffusion constant.
In contrast to the usual approach,\cite{Chalker} in which information
about the {\it anomalous diffusion exponent} $\eta$ is extracted from the
spectral function $S(q,\omega;E\approx E_c)$ we will be able to
extract information about the {\it localization exponent} $\nu$ using
the same spectral function but integrated over all energies $E$. %V0205
%J0203

We will show that $\Pi(q,\omega)$, which is an inherent 
fermionic disorder averaged two-particle
correlation function, can be re-expressed
as the {\it single particle} correlation function of 
an interacting (after the disorder average has been performed) 
dynamical system with an unusual action.
Therefore, in order to extract the dynamical 
behavior of the original problem, one simply has to study the disorder
averaged density of states of this new action. %J0203

The rest of this paper is organized as follows. In Sec.\ II
we introduce the model and mapping of the problem to 
the new ``Hamiltonian''. In Sec.\ III we calculate 
$\Pi(q,\omega)$ within the
self-consistent Born approximation. 
It displays normal diffusion at this level of
approximation. In Sec.\ IV we 
introduce the field theoretical approach to the disorder
averaging. 
In Sec.\ V we will demonstrate numerically the validity
of the scaling hypothesis stated in Eq.\ (\ref{argument}),
and finally in Sec.\ VI we present our conclusions.

%%%%%%%%%%%%%%%%%%%%%%%%%%%%%%%%%%%%%%%%%%%%%%%%%%%%%%%%%%%%%%%%%%%%%%%%%%

\section{Model and mapping}

We consider the two-dimensional spinless 
electron gas lying in the $x$-$y$ plane
which is subject to a perpendicular magnetic field ${\bf B} = B
\hat{\bf z}$ and an external potential $V({\bf r})$. $\hat{\bf z}$
denotes the unit vector in the $z$ direction.
In the symmetric gauge the vector potential is given by 
${\bf A} = -\frac{1}{2} {\bf r} \times {\bf B}$ and the  
one-particle Hamiltonian reads 
\begin{eqnarray}
H & = & H_0+H_D \nonumber \\
& = & \frac{1}{2m} \left[ {\bf p} + \frac{e}{c} {\bf A} \right]^2 +
V({\bf r}) \, .
\label{hamiltonian}
\end{eqnarray}
We restrict our investigations to the LLL and
thus project the Hamiltonian onto the states in the LLL. The
kinetic energy of all the LLL states is the same and after projecting
leads to a constant which we will neglect in what  follows. Writing  %J0206
the potential energy in Fourier space the Hamiltonian simplifies to
\begin{equation}
H=\sum_\qq v(-\qq )\bar{\rho}_\qq \,\,,
\label{simplifiedhamiltonian}
\end{equation}
where $v({\bf q})$ is the Fourier transform of the disorder potential.
The projected density operator is given by
\beq
\bar{\rho}_\qq  \equiv e^{-\frac{1}{4} \ell^2 q^2} \tau_\qq \,\,,
\eeq
with $\tau_\qq$ being the unitary magnetic translation operator which
translates the electron a distance 
$({\bf q} \times \hat{\bf z})\, \ell^2$.  
The formalism needed to project the
density operator $\rho_{\bf q} = e^{- i{\bf q}\cdot{\bf r}}$ onto the
LLL was developed elsewhere.\cite{girvin_jach} %V0210

The magnetic translation operators
have the following special property:
\beq
\tau_\qq \tau_\pp = \exp\left(\frac{i \ell^2}{2}q\wedge p\right) 
   \tau_{{\bf q}+{\bf p}} \,\,,
\eeq
where $ q \wedge p \equiv \left( {\bf q} \times {\bf p} \right) 
\cdot \hat{\bf z}$.
Hence their commutation relation defines a closed Lie algebra:
\beq
\left[\tau_\qq, \tau_\pp\right] = 2i 
\sin\left(\frac{\ell^2}{2}q\wedge p\right)
\tau_{{\bf q}+{\bf p}}\,\,.
\label{commutator}
\eeq
Also we have
\beq
{\rm Tr} \left\{ \tau_\qq \right\} = N \delta_{\qq,0}\, .
\eeq
The latter can be proved by noting that the left hand side 
is proportional to the one-particle 
trace of $\rbq$.  Since the trace is taken over states in
the LLL, the projection is unnecessary and we have
\beq
{\rm Tr} \left\{\rbq\right\} =
{\rm Tr} \left\{ e^{-i {\bf q}\cdot {\bf r}}   \right\}
\,\,,
\eeq
which vanishes unless ${\bf q}=0$.

If there are $N$ states in the Hilbert space, there are
$N^2$ independent operators on the space.  However there are exactly
$N^2$ different wave vectors on the torus, so the set of operators
$\rbq$ is ``complete'';  it spans the set of all operators.  The Hamiltonian
can be expressed in terms of the $\rbq$ and the Heisenberg 
equation of motion of the $\rbq$ is closed.  
This allows us to define the quantum ``Liouvillian'' matrix by
\beq
{\dot \tau_\qq}(t) = -i \sum_{\qq'} 
{\cal L}_{\qq \qq'} \tau_{\qq'}(t) \,\,.
\label{eqm}
\eeq
From the simple commutation properties Eq.\ (\ref{commutator})
of the $\tau_{\qq}$ it readily follows that
\beq
{\cal L}_{\qq \qq'} \equiv - \frac{2i}{\hbar}
v(\qq-\qq') e^{-\frac{1}{4} \ell^2 |\qq'-\qq|^2}
\sin\left(\frac{\ell^2}{2} \,q'\wedge q\right) \, .
\label{lioudef}
\eeq

Using the Liouvillian matrix 
we can immediately write down the formal solution of
the equation of motion Eq.\ (\ref{eqm}) for $\tau_\qq(t)$
\beq
\tau_\qq(t) = \sum_{\qq'} \left(e^{-i {\cal L} t}\right)_{\qq \qq'}
\tau_{\qq'}(0) \, .
\eeq 
This leads to a simple expression for the density-density correlation
function defined in Eq.\ (\ref{pidef})
\beq
\tilde\Pi(q,t) = -i \frac{\theta(t)}{\hbar \ell^2}
e^{-\frac{1}{2} \ell^2 q^2}\left< \! \left<
\left( e^{-i {\cal L} t} \right)_{\qq \qq} \right> \! \right> \, .
\eeq
We can define an $N^2$ element operator ``superspace'' and view ${\cal
  L}$
as the ``Hamiltonian''.  From this point of view, finding
${\rm Im}\, \Pi(q,\omega)$ is the same as finding the
{\it one-particle density of states} 
for a system with Hamiltonian  ${\cal L}$:
\begin{eqnarray}
\Pi(q,\omega) & = & -\frac{i}{\hbar \ell^2}
e^{-\frac{1}{2} \ell^2 q^2}
\int_0^\infty dt
e^{i(\omega + i\delta)t} \left< \! \left<  \, \left< \qq \left| 
e^{-i {\cal L} t} \right| \qq \right>
\, \right> \! \right> \nonumber \\
& = & \frac{1}{\hbar \ell^2}  e^{-\frac{1}{2} \ell^2 q^2}
\left< \! \left<  \, \Bigl< \qq \Bigr| 
\frac{1}{\omega + i\delta - {\cal L}} 
\Bigl|  \qq \Bigr> \, \right> \! \right> \, ,
\label{eq:A}
\end{eqnarray}
where we have introduced states $\left| \qq \right>$ with 
$\left< \qq \right| {\cal L} \left| \qq' \right> 
\equiv {\cal L}_{\qq \qq'} $ and $\delta$ is an infinitesimal small
positive number. %S0209

This remarkable formula is our central result. Let us now try to understand
its import. In a crude sense it represents a kind of bosonization of the
problem. Ordinarily in an interacting many-body system the equations 
of motion for the density are not closed but rather involve a 
hierarchy of additional operators. However for the special case 
of one-dimension and a linear dispersion 
relation (the Tomonaga-Luttinger model) the equations of motion {\em are}
closed and the density fluctuations become free bosons\cite{Fradkin} even
though the underlying particles are interacting. In the present problem
(without electron-electron interactions) the equations for the 
density operators
close after projection onto a single Landau level 
(which for simplicity we have taken to be the lowest). This has 
several advantages. First we do not have to work separately with 
retarded and advanced one-particle Green's functions and their
products.  %V0210
Secondly we note that there are no problems with gauge invariance and 
conserving approximations. This is because the Liouvillian matrix 
elements ${\cal L}_{\qq \qq'}$ vanish if either $\qq$ or $\qq'$ 
vanish. Thus the total charge in the system is automatically 
conserved. Finally this representation allows
us to establish a hierarchy of length and time scales which 
should be suitable for renormalization group (RG) analysis. Because 
the kinetic energy has been quenched, high momentum of a particle 
is not associated with high energy. Since the Liouvillian vanishes 
at small wavevectors, it naturally organizes the decay rates of
density fluctuations into short time scales at large wavevectors 
and long time scales at small wavevectors. As we comment further 
below however, there are technical obstacles to be overcome before 
this RG can be carried out.

We take the disorder to be gaussian distributed, but
not necessarily white noise, i.e.\ possibly smoothed.  We then have
\beq
\langle\!\langle v({\bf q}) \rangle\!\rangle = 0 \, 
\eeq
and 
\beq
\langle\!\langle v({\bf q})v({\bf q}') \rangle\!\rangle
= \frac{2\pi  \alpha^2 v^2}{L^2} e^{-\frac{1}{2} \ell^2 
q^2(\alpha^2 -1)} \delta_{{\bf q}+{\bf q}',0}\,\,,
\label{eq:disorder_variance}
\eeq
which in real space translates into
\beq
\langle\!\langle V({\bf r})V({\bf r}') \rangle\!\rangle=
\frac{\alpha^2 v^2}{\ell^2 (\alpha^2-1)}
\exp{\left[ -\frac{|{\bf r}-{\bf r}'|^2}{2 \ell^2 (\alpha^2-1)} \right] }\,.
\eeq
Here $v$ denotes the strength of the disorder potential and   %J0206
$\alpha$ is a dimensionless 
smoothness parameter. In the limit of a distribution which is
extremely smooth ($\alpha \to \infty$),
the one-particle electronic density of states approaches a 
gaussian\cite{Raikh}
\beq
\rho_{\alpha=\infty}(E) = \frac{1}{(2\pi)^{3/2} \ell^2 v} 
\exp\left[ -\frac{1}{2 v^2}(E-E_c)^2 \right] \, .
\label{doslargealpha}
\eeq
%J0203
An integration over all energies $E$ gives the number of
states in the LLL divided by the sample area $N/L^2$ which 
is $1/(2 \pi \ell^2)$. 
For $\alpha = 1$ the disorder distribution goes over 
to the uncorrelated white noise distribution for which Wegner has 
determined the density of states.\cite{Wegner} At $E=E_c$ it is given
by
\beq
\rho_{\alpha=1}(E_c) = \frac{\sqrt{2}}{\pi^2 \ell^2 v} \, .
\label{dosalpha1}
\eeq 

%%%%%%%%%%%%%%%%%%%%%%%%%%%%%%%%%%%%%%%%%%%%%%%%%%%%%%%%%%%%%%%%%%%%%%%%%%%%

\section{Self-consistent Born approximation}

\begin{figure}[htb]
\epsfxsize=3.375in
\centerline{\epsffile{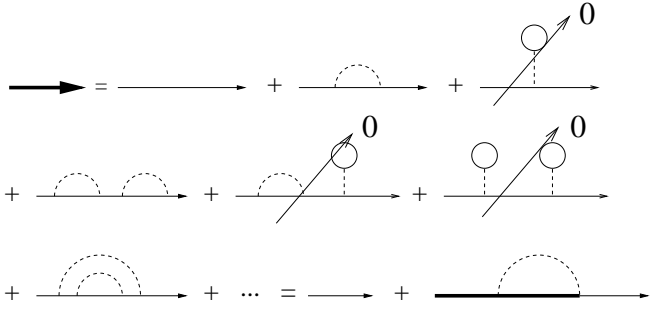}}
\vspace{0.3cm}
\caption{Partial sum of all non-crossing diagrams of the propagator
  $\hat \Pi(q,\omega)$. For details see the text.}
\label{fig:SCBA}
\end{figure}

We next calculate $\Pi(q,\omega)$ Eq.\ (\ref{eq:A}) 
in the self-consistent Born approximation. We define 
the complex self-energy 
$\Sigma(q,\omega)=\Sigma_R(q,\omega) + i \Sigma_I(q,\omega)$ 
for the propagator 
\beq
\hat \Pi(q,\omega) \equiv \hbar \ell^2
e^{\frac{1}{2} \ell^2 q^2} \Pi(q,\omega)
\,\, .
\label{propdef}
\eeq
by setting 
\beq
\hat \Pi(q,\omega)=\frac{1}
{\omega+i\delta -\Sigma(q,\omega)}
\,\, .
\label{selfenergy}
\eeq
Within the self-consistent Born approximation 
the self-energy is given by the expression
\begin{eqnarray}
\Sigma^{B}(q,\omega) =  \sum_\pp
\left< \! \left< {\cal L}_{\qq,\qq+\pp} {\cal L}_{\qq+\pp,\qq} 
\right> \! \right>  \hat \Pi^{B}(| \qq + \pp |,\omega)\,\,.
\label{scba}
\end{eqnarray}
In contrast to standard many-body perturbation theory 
the right hand side of this expression does not contain an energy sum.  
In this approximation all non-crossing diagrams for the propagator 
$\hat \Pi(q,\omega)$ are summed, as shown in Fig.\ \ref{fig:SCBA}. 
In this figure a thick solid line stands for  
$\hat \Pi^{B}(q,\omega)$ and a thin solid line indicates  
the ``noninteracting'' propagator $\hat \Pi^{0}(q,\omega)$, 
which is given by Eq.\ (\ref{selfenergy}) 
with $\Sigma(q,\omega) \equiv 0$. $\hat \Pi^{0}$ 
is independent of $q$. %J0203
The consequences of this for a perturbative treatment will be
discussed in the next section. %V0205
In Fig.\ \ref{fig:SCBA} the vertex with an 
incoming and an outgoing solid line and a dashed line stands for a
matrix element ${\cal L}_{\qq \qq'}$ of the Liouvillian. 
The disorder average introduces ``contractions'', i.e.\ connections,
between the dashed lines.
In general the
Hartree terms are included in the partial sum, but as indicated in
Fig.\ \ref{fig:SCBA}, they vanish because of the $q\wedge p$ term 
in the matrix elements of the Liouvillian.

Using the distribution introduced in the last section [see Eq.\
(\ref{eq:disorder_variance})] and the definition of the Liouvillian
matrix Eq.\ (\ref{lioudef}) we obtain the self-consistency equation  
\beq
\Sigma^B(q,\omega)=\frac{ 2\pi \alpha^2 v^2}{\hbar^2 L^2} \sum_\pp
\frac{e^{-\frac{1}{2} \ell^2 \alpha^2|\qq-\pp|^2}
4\sin^2(\frac{\ell^2}{2}q\wedge p)}{\omega+i\delta -\Sigma^B(p,\omega)}\,\,.
\label{self_energy}
\eeq
The strength of the disorder $v$ can be scaled out of this equation by
replacing $\Sigma^B \to \hbar \Sigma^B/v$ and $\omega \to \hbar
\omega/v$.

\begin{figure}
\epsfxsize=2.60in
\centerline{\epsffile{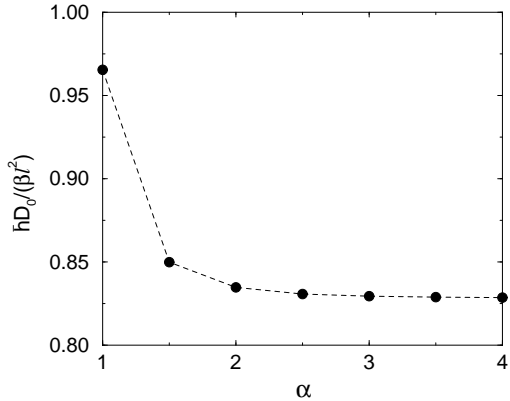}}
\caption{Diffusion constant $D_0$ as a function of the smoothness of
  the disorder $\alpha$.}
\label{Dfig}
\end{figure}

As explained in the introduction the diffusive properties can be read
off from the small $q$ and $\omega$ limit of the 
imaginary part of $\Pi$. 
For $q \to 0$ we have $\hat \Pi(q,\omega) \to
\hbar \ell^2 \Pi(q,\omega)$ and can thus write
\begin{eqnarray}
\hbar \ell^2 \, {\rm Im} \, \Pi^B(q,\omega) = \frac{\Sigma_I^B(q,\omega)}{\left[ \omega 
- \Sigma_R^B(q,\omega) \right]^2 +\left[ \Sigma_I^B(q,\omega) \right]^2} . 
\label{impi}
\end{eqnarray}
Eq.\ (\ref{self_energy}) can be solved
numerically by iteration. Following Eq.\ (\ref{argument})   %J0203
the best way to extract the
diffusive properties is a ``scaling'' plot in which (for fixed $v$
and $\alpha$)  
$- \hbar \omega \ell^2 {\rm Im} \, \Pi^B(q,\omega)$ is plotted as a
function of $q L_{\omega}$ for different small $q$ and $\omega$. 
Such an evaluation shows that on this level of
approximation $ - \hbar \omega \ell^2 {\rm Im} \, \Pi^B(q,\omega)$ 
is a function of  $q L_{\omega}$  only and thus does not display  
a sign of the prefactor $(q \xi_0)^{1/\nu}$ discussed in connection
with Eq.\ (\ref{argument}). Furthermore ${\rm Im} \, \Pi^B(q,\omega)$
only shows {\it normal diffusion} with a diffusion constant $D_0$ which 
for $q \to 0$ and $\omega \to 0$ is independent of $q L_{\omega}$. 
We thus conclude that (as expected) the occurrence of the critical 
exponents $\nu$ and $\eta$ is a {\it higher order fluctuation effect.}  %S0209
For $\omega \to 0$,
$\Sigma_R^B(q,\omega)$ goes to zero for all $q$. Thus $D_0$ is given by
\begin{eqnarray}
D_0 = - \lim_{\omega \to 0} \lim_{q \to 0}  
\Sigma_I^B(q,\omega) / q^2 \, . 
\label{extract}
\end{eqnarray}
Because of the scaling property discussed following Eq.\
(\ref{self_energy}) $D_0$ is proportional to $v$. As shown 
in Fig.\ \ref{Dfig} $D_0$ also depends on the smoothness $\alpha$
of the disorder. Between $\alpha=1$ (white noise) and $\alpha=2$,  
$D_0$ changes by approximately ten percent. For $\alpha>2$ the
$\alpha$ dependence is extremely weak and for $\alpha \to \infty$, 
$D_0$ saturates at $D_0^{\alpha=\infty} \approx  0.828 \, 
v \ell^2/\hbar$. 
For $\alpha=1$ we find $D_0^{\alpha=1} \approx 0.965\, v
\ell^2/\hbar $.

Using the Einstein relation for the conductivity %J0203
and Eqs.\ (\ref{doslargealpha}) and  
(\ref{dosalpha1}) we obtain
\begin{eqnarray}
\sigma_{ \rm xx}^{\alpha=\infty} \approx 0.330 \,  \frac{e^2}{h}
\label{galphainf}
\end{eqnarray}
and
\begin{eqnarray}
\sigma_{ \rm xx}^{\alpha=1} \approx 0.869 \,  \frac{e^2}{h} \, .
\label{galpha1}
\end{eqnarray}
If one is interested in the large $\alpha$ limit it might be tempting
to expand the sine in Eq.\ (\ref{self_energy}), as only small $p$ 
contribute to the sum due to the exponential function. 
Anticipating that for small $q$ the self-energy is quadratic in $q$ the
ansatz $\Sigma^B(q,\omega) = - i q^2 \tilde D_0$ seems to be
plausible. 
Then the self-consistency equation can be solved analytically leading to 
$\tilde D_0 = (1/ \sqrt{2}) v \ell^2/\hbar \approx 0.707 
v \ell^2/\hbar$. A comparison with $D_0^{\alpha=\infty}$ discussed
above shows that this procedure {\it does not} give the correct large
$\alpha$ value for $D_0$. This is due to the fact that in the exact %J0203
solution of Eq.\ (\ref{self_energy}) the range of $q$ values over
which $\Sigma^B(q,\omega)$ can be approximated by a purely quadratic
function in $q$ shrinks as $1/\alpha$. Thus in
the limit $\alpha \to \infty$ it would be necessary to include 
higher order terms
in the expansion of $\Sigma^B(q,\omega)$ in order to reproduce
the numerical result in Eq.\ (\ref{galphainf}).  %J0203
Note that $\sigma_{ \rm xx}^{\alpha\gg 1}$ obtained above 
is independent  %V0205
of the correlation parameter
$\alpha$ as it should in the limit $\alpha \gg 1$. Since the 
exact conductivity is universal, the present result is a considerable
improvement over the traditional self-consistent Born approximation
result for which the conductivity vanishes like $\alpha^{-1}$ in this
limit.\cite{Ando}  %S0209

In a previous numerical study\cite{huoandbhatt2} it was found that 
$\sigma_{ \rm xx}= (0.54 \pm 0.04) e^2/h$ independent of the
smoothness of the disorder. The results for $\sigma_{ \rm xx}$
obtained within our approach are of
the same order of magnitude as the one calculated using purely
numerical methods\cite{huoandbhatt2} but in contrast to this one  
our results depend on $\alpha$. This is
due to the fact that we have calculated $D_0$ within the 
self-consistent Born approximation but included in the
Einstein relation the {\it exact} density of states at the critical energy.

Using our approach of calculating the disorder averaged 
one-particle
correlation function for the dynamical system described by the
Liouvillian, we observe normal
diffusion already at the level of the self-consistent Born approximation.
In the usual fermionic picture of noninteracting
electrons in the presence of disorder and a magnetic field,
much more elaborate techniques, as e.g.\ Borel resummation, instanton
methods, the replica trick, and the supersymmetry method, are 
used to obtain similar results.\cite{Hajdu book} 
In particular, in the more traditional approaches, diffusion is not obtained
at the saddle point level and it is necessary to include gaussian fluctuations
(i.e. sum ladder diagrams) to obtain diffusion. Because we deal directly
with the density itself, we obtain diffusion even at the saddle point level. %S0209

%%%%%%%%%%%%%%%%%%%%%%%%%%%%%%%%%%%%%%%%%%%%%%%%%%%%%%%%%%%%%%%%%%%%%%%%%%

\section{Field theoretical approach}

To go beyond the self-consistent Born approximation it might prove
advantageous to bring our approach into a field theoretical
framework. This is what we will do in this section. 
In reformulating $\Pi(q,\omega)$ using field theoretical
methods we use the gaussian integral identity
\begin{equation}
-i \left\langle\bar\psi_{\bf q}
\psi_{\bf q}\right\rangle  = 
\Bigl< \qq \Bigr| 
\frac{1}{\omega + i\delta - {\cal L}} 
\Bigl|  \qq \Bigr>  \, ,
\label{gaussid}
\end{equation}
where  
\begin{equation}
\left\langle\bar\psi_{\bf q} \psi_{\bf q}\right\rangle
\equiv \frac{1}{Z}\; \int {\cal D}\bar\psi {\cal D}\psi\; e^{-S_{\psi}}\;
\bar\psi_{\bf q} \psi_{\bf q}\,\,,
\end{equation}
and
\begin{equation}
S_{\psi} 
\equiv -i \sum_{{\bf k},{\bf k}'} \bar\psi_{\bf k}\; [\omega + i\delta -
{\cal L}]_{{\bf k},{\bf k}'}\; \psi_{{\bf k}'}. 
\end{equation}
The $\psi_{\bf q}$ denote complex (bosonic) fields and $Z$ is given by 
\begin{equation}
Z \equiv \int {\cal D}\bar\psi {\cal D}\psi\; e^{-S_{\psi}} \,\,.
\end{equation}
In order to ensemble average over the disorder we introduce 
additional {\it Grassmann variables} to represent $1/Z$ as a path
integral\cite{Hajdu book} 
\begin{equation}
\frac{1}{Z} = \int {\cal D}\bar\xi{\cal D}\xi\, e^{-S_\xi},
\end{equation}
where
\begin{equation}
S_\xi \equiv -i \sum_{{\bf k},{\bf k}'} \bar\xi_{\bf k}\; [\omega + i
\tilde\delta -
{\cal L}]_{{\bf k},{\bf k}'}\; \xi_{{\bf k}'}. 
\end{equation}
One can then carry out the ensemble average over the gaussian
distributed disorder and obtains the generalized functional
\begin{equation}
\bar Z(\omega) =
\int {\cal D}{\bar\xi} {\cal D}\xi
\int {\cal D}{\bar\psi} {\cal D}\psi
\; e^{-F(\omega)} \,\,,
\end{equation}
where
\begin{eqnarray}
F(\omega) & \equiv& \sum_{\bf k} \left[(-i\omega +\delta_{\bf k})\;
\bar\psi_{\bf k} \psi_{\bf k}
+(-i\omega+\tilde\delta_{\bf k})\:
\bar\xi_{\bf k} \xi_{\bf k} \right]\nonumber\\
&&+ \sum_{{\bf k},{\bf k}'} \sum_{{\bf p},{\bf p}'}
\left< \! \left< 
{\cal L}_{{\bf kk}'}{\cal L} _{{\bf pp}'}  \right> \! \right> \;
\left[ 
\bar\psi_{\bf k} \bar\psi_{\bf p} \psi_{{\bf p}'} \psi_{{\bf k}'}
\right. \nonumber \\
&& \left. +2\bar\psi_{\bf k} \psi_{{\bf k}'} \bar\xi_{\bf p} \xi_{{\bf p}'} 
+ \bar\xi_{\bf k} \bar\xi_{\bf p} \xi_{{\bf p}'} \xi_{{\bf k}'}
\right] \, .
\end{eqnarray}
Here we have  let $\delta \rightarrow \delta_{\bf k}$ so that
we can generate the correlation functions by
\begin{equation}
\left\langle\bar\xi_{\bf q} \xi_{\bf q}\right\rangle_{\omega} =
\left\langle\bar\psi_{\bf q} \psi_{\bf q}\right\rangle_{\omega} =
-\frac{\partial\, \bar{Z}(\omega)}{\partial\, \tilde\delta_{\bf q}}
= -\frac{\partial\, \bar{Z}(\omega)}{\partial\, \delta_{\bf q}}. 
\end{equation}
Once the disorder averaging is done we finally obtain
\begin{eqnarray}
F(\omega) & = & -i\sum_\qq \left[
(\omega+i \delta_\qq) \bar \psi_\qq \psi_\qq+
(\omega+i \tilde\delta_\qq)\bar\xi_{\bf q} \xi_{\bf q} 
\right] \nonumber \\
&&+\sum_{\qq_1,\qq_2,\qq_3,\qq_4} f(1,2,3,4) \left[ 
\bar\psi_{\qq_1} \bar\psi_{\qq_2}
\psi_{\qq_3}\psi_{\qq_4} \right. \nonumber\\
&&\left. + 2 \bar\psi_{\qq_1}\psi_{\qq_4} \bar\xi_{\qq_2}\xi_{\qq_3}
+\bar\xi_{\qq_1} \bar\xi_{\qq_2}\xi_{\qq_3}\xi_{\qq_4} \right] \,,
\label{finalaction}
\end{eqnarray}
with
\begin{eqnarray}
f(1,2,3,4)& = & - \frac{\pi \alpha^2 v^2}{\hbar^2 L^2} 
e^{-\frac{1}{2} \ell^2 \alpha^2 |\qq_1-\qq_4|^2} 
4 \sin{\left( \frac{\ell^2}{2} q_1 \wedge q_4 \right)} 
\nonumber\\ 
&&\times\sin{\left( \frac{\ell^2}{2} q_2 \wedge q_3 \right) } \,
\,\delta_{\qq_1+\qq_2,\qq_3+\qq_4}\,.
\label{intera}
\end{eqnarray}

In contrast to standard many-body theory the action 
Eq.\ (\ref{finalaction}) does
{\it not} contain a sum over the frequency. $\omega$ only enters this
equation as an {\it external parameter}. As already 
discussed in the last section  the noninteracting
propagator ($v=0$) is given by $(\omega + i \delta)^{-1}$ and 
does not depend on $q$. Thus a perturbation theory or RG
procedure can only be set up after a $q$ dependent propagator 
has been generated by self-consistently summing up an {\it entire
  class} of diagrams, as e.g.\ the non-crossing  diagrams in Sec.\
III. Furthermore the interaction $f$ in Eq.\ (\ref{intera}) has an
unusual momentum dependence compared to standard standard $\phi^4$
theory of critical phenomena: It {\it vanishes} if one of the ${\bf q}_i$ 
goes to zero and is {\it periodic} in the momenta. 

Using the field theoretical approach 
we can reproduce the approximation discussed in Sec.\ III, which is
usually called self-consistent mean-field or saddle point approximation in
the present context.  
In the absence of symmetry breaking, the middle of the three 
quartic terms in the action 
cannot contribute to the saddle point solution since its 
coefficient vanishes for $\qq_1=\qq_4$ and $\qq_2=\qq_3$. 
Hence we can deal separately with the
bosonic and the fermionic variables when discussing 
the saddle point solution.
By performing the usual pairing of the fields in the quartic interaction
term  at the mean-field level we have  %J0206
\begin{eqnarray}
\bar\psi_{\qq_1}\bar\psi_{\qq_2}\psi_{\qq_3}\psi_{\qq_4} & = & 
i\hat \Pi^{\rm MF}(q_1,\omega) 
\delta_{\qq_1,\qq_3} \bar\psi_{\qq_2}\psi_{\qq_4}
\nonumber \\
&& +i \hat \Pi^{\rm MF}(q_2,\omega) 
\delta_{\qq_2,\qq_4} \bar\psi_{\qq_1} \psi_{\qq_3}  \,  .
\end{eqnarray}
Thus we can write
\beq
F^{\rm MF}(\omega) = \sum_\qq \; \bar\psi_\qq 
\left[ -i\omega+\delta+i\Sigma^{\rm MF}(q,\omega) \right] \psi_\qq\,\,,
\eeq
and use this in calculating 
\begin{eqnarray} 
i \hat \Pi^{\rm MF}(q,\omega) & \equiv &  
\langle\bar\psi_\qq\psi_\qq\rangle_{\omega}^{\rm MF }
 = {\int{\cal D} \bar\psi {\cal D} 
\psi e^{-F^{\rm MF}(\omega)}\bar\psi_\qq\psi_\qq
\over \int{\cal D} \bar\psi {\cal D} \psi e^{-F^{\rm MF}(\omega)}}
\nonumber \\ 
& = & {i\over \omega+i \delta -\Sigma^{\rm MF}(q,\omega)} = 
i \hat \Pi^{\rm B}(q,\omega) \, ,
\end{eqnarray}
which reproduces the self-consistency Eq.\ (\ref{self_energy}) for the
self-energy.

At present we do not know how to evaluate the correlation function 
beyond the self-consistent mean-field approximation in a controlled way. 
However, we hope that in the future it will be possible to analytically
extend our results.

%%%%%%%%%%%%%%%%%%%%%%%%%%%%%%%%%%%%%%%%%%%%%%%%%%%%%%%%%%%%%%%%%%%%%%%%%%%%

\section{Numerical results}

In this section we will numerically calculate 
${\rm Im} \, \Pi(q,\omega)$ by exact diagonalization and 
verify the scaling hypothesis stated in Eq.\ (\ref{argument}).  
We closely follow the procedure and
notation used by Chalker and Daniell.\cite{Chalker} 
Motivated by Eq.\ (\ref{relation}) we define
\begin{eqnarray}
&& \bar S(r,\omega)  \equiv  -\frac{1}{2 \pi^2} \, {\rm Im}\, \bar 
\Pi(r,\omega) \nonumber \\
& & =
\Bigl< \! \Bigl<  
\sum_{i,j} \delta(\hbar \omega + E_i - E_j) 
\psi_i(0)\psi_i^*({\bf r})\psi_j({\bf r})\psi_j^*(0)
\Bigr> \! \Bigr>  \, .
\label{S}
\end{eqnarray}
The single particle wave functions $\psi_i({\bf r})$ can be expanded
in the basis of the elliptical theta
functions $\phi_{m}({\bf r})$ 
\begin{eqnarray}
\psi_i({\bf r}) = \sum_{m=1}^{N} a_i(m) \phi_m({\bf r}) \, ,
\label{expansion}
\end{eqnarray}
where
\begin{eqnarray}
\phi_{m}(x,y) & = &\frac{1}{\sqrt{L \ell \pi^{1/2}}}
\sum_{s=-\infty}^\infty \exp{ \left( i X_{m,s} y / \ell^2 \right)}
\nonumber \\
&& \times \exp{\left[  - (x- X_{m,s})^2 / (2 \ell^2) \right] }\, ,
\label{thetafundef}
\end{eqnarray}
and 
\begin{eqnarray}
X_{m,s}= m \frac{2\pi}{L} \ell^2 +  s L \, .
\label{Xdef}
\end{eqnarray}
Then the Fourier transform of Eq.\ (\ref{S}) can be written as
\begin{eqnarray}
S(q,\omega) & = & \frac{1}{2 \pi \ell^2 N^2}e^{-\frac{1}{2} \ell^2 q^2}
\nonumber \\
&& \times \left< \! \left<  
\sum_{i,j} \delta(\hbar \omega+E_i-E_j) \, 
Q_{i j}(k,l)
\right> \! \right>\,,
\label{SQ}
\end{eqnarray}
where   
\begin{eqnarray}
Q_{i j}(k,l) & = & N \Biggl| \sum_{m=1}^{N} a_i(m) 
a_j^{\ast}([m-l]) \nonumber \\ 
&& \times \exp{ \left( i 2 \pi k \frac{m}{N} \right) }
\Biggr|^2 \, ,
\label{Qdef}
\end{eqnarray}
and ${\bf q} = (2 \pi /L ) (k,l) = \sqrt{\frac{2 \pi }{\ell^2 N}}
(k,l)$, with $k$, $l$ integer.
In Eq.\ (\ref{Qdef}) $[m+l]$ is defined as being
$m+l$ for $1\le m+l\le N$ and $m+l\pm N$ otherwise such that 
$1\le |m+l\pm N|\le N$.
In the numerical calculation we replace 
the delta function in Eq.\ (\ref{SQ}) 
by a sharply peaked gaussian $ \delta_{\gamma}(x) \propto 
\exp[-x^2/\gamma^2]$ with a broadening\cite{Chalker} 
$\gamma = 0.64 \, v / N$ which is of the order of the level spacing. 
We then have
\begin{equation}
S(q,\omega)= \frac{1}{2 \pi \ell^2 N^2}e^{-\frac{1}{2} \ell^2 q^2}
K(q,\omega)\,,
\label{SK}
\end{equation}
with
\begin{equation}
K(q,\omega)=\frac{\left< \! \left< 
\sum_{i \neq j} \delta_{\gamma}( \hbar \omega+E_i-E_j)
Q_{ij}(k,l) \right>  \! \right> }
{\left< \! \left<  \sum_{i \neq j}  \delta_{\gamma}(\hbar \omega+E_i-E_j) 
\right> \! \right> }\,.
\end{equation}

This function is suitable for a numeri\-cal in\-vestiga\-tion.\cite{Chalker} 
We restrict ourselves to a white noise disorder distribution
($\alpha=1$).  
We calculate $K(q,\omega)$ for values of
$2 \leq k^2+l^2 \leq 25$ and $\hbar \omega= \gamma n$, with 
$3 \leq n \leq 23$, where 
the limits have been chosen such that $L^{-1}< q < \ell^{-1}$ 
and $\hbar \omega \ll v$ but $\hbar \omega$ greater
than the level spacing of the finite size system. 
The system sizes range from $N=200$ to $N=2000$, and the 
number of disorder realizations are 500 or 100 depending on the 
system size. All values of $K(q,\omega)$ were determined
to an accuracy better than 1\% in the disorder averaging.

\begin{figure}
\epsfxsize=2.60in
\centerline{\epsffile{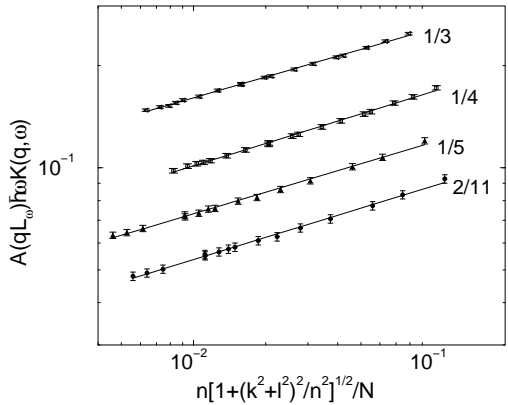}}   %J0203  %V0205
\caption{$\hbar \omega K(q,\omega)$ for different fixed 
$(q L_\omega)^2\propto(k^2+l^2)/n$ and 
small $q$ and $\omega$ as a function 
of $n [1+(k^2+l^2)^2/n^2]^{(1/2)}/N$ on a log-log scale.
For clear comparison each data set has
been multiplied by a constant 
factor $A(q L_\omega)$ and is labeled by the ratio $(k^2+l^2)/n$.}
\label{scal}
\end{figure}

For a fixed and small value of $q L_\omega$  
(so that we are in the range of  %J0203
normal diffusion) % given by \cite{Chalker} $q L_\omega<7.74$)  %V0205 
and $q,\omega \rightarrow 0$ 
we expect from Eq.\ (\ref{argument}) that $\hbar \omega K(q,\omega)$ 
scales as $(\hbar \omega/ v )^{\frac{1}{2\nu}} \propto 
(q \ell)^{\frac{1}{\nu}}$. 
The scaling hypothesis is illustrated in Fig.\ \ref{scal}, where we plot 
$A(q L_\omega) \hbar \omega K(q,\omega)$ for fixed ratios of 
$(q L_\omega)^2 \propto (k^2+l^2)/n$ as a
function of $n \sqrt{1 + (k^2+l^2)^2/n^2}/N$ on a log-log scale. %J0203 %V0205
Here each curve is multiplied by a constant factor $A(q L_\omega)$
[different for each $(k^2+l^2)/n$ ratio] to make the comparison 
of the different lines easier. Also the factor 
$\sqrt{1 + (k^2+l^2)^2/n^2}$ multiplying $n/N \propto \omega$ is %V0205
used such that the curves line up horizontally. %S0209
The fact that data calculated for different system sizes fall
onto the same curve [for a fixed ratio of $(k^2+l^2)/n$)]
indicates that the limits chosen above for $k,l$, and $n$
do avoid large finite size effects. 
On the log-log scale the
different data sets fall onto straight lines and can be fitted
by power-laws (solid lines in Fig.\ \ref{scal}).  

\begin{figure}
\epsfxsize=2.60in
\centerline{\epsffile{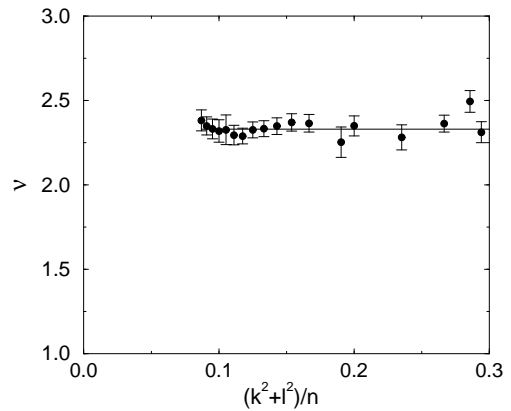}}   %J0203 %V0205
\caption{Localization exponent $\nu$ obtained
from Fig.\ 3. The solid line corresponds to the average $\nu$.}
\label{nu}
\end{figure}

The localization exponent $\nu$ extracted from 
the slope of the lines in Fig.\ \ref{scal} is shown as a function 
of $(k^2+l^2)/n \propto (q L_\omega)^2$ in Fig.\ \ref{nu}. %V0205
Within our error bars and for the $q L_\omega $
considered,  $\nu$ is a constant. 
Its value $\nu=2.33 \pm 0.05$ is in excellent agreement with
previous finite-size scaling 
studies\cite{Hajdu book,Huckestein RMP,Huckesteinkramer,Mieck,huo_and_bhat}
and strongly supports the scaling hypothesis Eq.\ (\ref{argument}).
The fact that the lowest $(k^2+l^2)/n$ points seem to be moving upwards
in Fig.\ \ref{nu} 
is an indication that there are still some finite-size 
effects for the low values of $(k^2+l^2)$.  %J0209 %V0210
In contrast to previous numerical 
studies we are able to obtain
information about the critical exponent $\nu$ from systems of finite size
without doing finite-size scaling.  %J0203

\section{Conclusion}
We have presented a new analytical and numerical approach to the
localization-delocalization transition in the LLL of the 
IQHE. 
By using the closed 
Lie algebra of the density operators in the LLL we are able to write 
the equation of motion for the densities in a  closed form which  %J0203
can be solved formally. 
Using the solution of the
equation of motion for the projected densities we can express the integrated
spectral function %J0203
$\int dE S(q,\omega;E)$ as the disorder averaged density of states of 
a dynamical system with a novel action. We show analytically 
that the self-consistent mean-field approximation of the 
integrated spectral function yields normal diffusion but it 
misses the critical scaling. %J0203
However, it is encouraging to note that even
at this level of approximation the longitudinal conductivity 
is in approximate agreement with previous numerical 
studies.\cite{huoandbhatt2} %J0203 %V0205
Finally, using exact diagonalization, we are able to extract 
the localization critical exponent $\nu$ from the 
integrated spectral function by using 
the scaling hypothesis Eq.\ (\ref{argument}), without having to do 
finite-size scaling. We obtain $\nu=2.33\pm0.05$ in excellent agreement
with previous studies.\cite{Huckesteinkramer,Mieck,huo_and_bhat} %V0205
We hope that in the future it will be possible
to extend our approach beyond the self-consistent
mean-field level and analytically extract information  %J0203
about the critical exponent $\nu$. %J0203

\vspace{0.4cm}

The authors would like to thank A.\ Zee, A.H.\ MacDonald, T.\ Brandes,
K.\ Sch\"onhammer, and J.T.\ Chalker for helpful discussions. 
V.M.\ is grateful to the Deutsche Forschungsgemeinschaft for 
financial support during his stay at Indiana University. 
This work was furthermore supported by the NSF Grant No.\ DMR-9714055
and the NSF Grant No.\ DMR-9820816.

\end{document}